\documentclass[prd,showpacs,twocolumn]{revtex4}
\usepackage{epsfig,dcolumn,graphicx,bm}

\def\x{{\bf x}}
\def\y{{\bf y}}
\def\k{{\bf k}}
\def\q{{\bf q}}
\def\p{{\bf p}}
\def\ds{\displaystyle}
\def\vp{\varphi}
\def\too{\mathop{\to}\limits_{n\to\infty}}

\begin{document}

\title{Microscopic derivation of the pion coupling to heavy--light mesons}

\author{A.V. Nefediev}
\affiliation{Institute of Theoretical and Experimental Physics, 117218,
B.Cheremushkinskaya 25, Moscow, Russia}

\author{J.E.F.T. Ribeiro}
\affiliation{Centro de F\'\i sica das Interac\c c\~oes Fundamentais
(CFIF),Departamento de F\'\i sica, Instituto Superior T\'ecnico, Av.
Rovisco Pais, 1049-001 Lisbon, Portugal}

\author{Adam P. Szczepaniak}
\affiliation{Department of Physics and Nuclear Theory Center, \\ Indiana University,
Bloomington, IN 47405, USA }

\begin{abstract}
The  Goldberger--Treiman relation for heavy--light systems is
derived in the context of a quark model. As a paradigmatic
example, the case of ${\bar D}\to {\bar D}' \pi $ is studied in detail. The
fundamental role played by the pion two-component wave function,
in the context of the Salpeter equation, is emphasized.
\end{abstract}


\maketitle
\section{Introduction}

Quark models endowed with effective quark current-current
microscopic interactions, consistent with the requirements of
chiral symmetry, have provided, through the years, an invaluably
useful tool towards the construction of an effective low--energy
theory for the strong interactions
\cite{Orsay,Orsay2,Lisbon,linear,ASES}. Besides regularizing the
ultraviolet divergences of the theory, these effective theories
bring in the necessary interaction scale needed to make contact
with the hadronic phenomenology. An important feature of this
class of models is the essential convergence of results and
conclusions across a variety of possible forms for the confining
kernel. They fulfill the well--known low energy theorems of
Gell--Mann, Oakes, and Renner \cite{Orsay2}, Goldberger and
Treiman \cite{Lisbon}, the Weinberg theorem \cite{EmilCota}, and
so on. Using this formalism it is also possible to give, for this
class of models, an analytic proof of the low--energy theorems in
the light--quark sector \cite{BicudAp}. It turns out that the
chiral angle --- the solution to the mass-gap equation--- remains
the only nontrivial characteristic of such a class of models and
it defines the latter completely.

In the present paper, we derive the generalized
Gold\-berger--Treiman (GT) relation for the heavy--light systems.
The heavy--light mesons have received a lot of attention in the
past few years due to the discovery of new narrow states in the
$c{\bar s}$ family~\cite{ds}. Nowak et al.~\cite{Nowak} and
Bardeen and Hill~\cite{BH} have postulated that the spectrum of
such states should reflect the pattern of dynamical chiral
symmetry breaking. Namely in the heavy--quark limit, the mass
spectrum is expected to be determined by the light quark and, were
chiral symmetry to be exact, the heavy--light states of opposite
parity would have become degenerate. It then follows that physical
splitting between parity doublers could be related to the scale of
spontaneous breaking of chiral symmetry, that is the quark condensate
or, phenomenologically, to the constituent quark mass. As more
states in the open charm sector have been reported, alternative
pictures have been examined, making this sector a good testing
ground for phenomenological models~\cite{others}. Recently, for
example, a canonical, quark model description has been
investigated and it was argued that the new $D_s$ states can in
fact be described as quark model states~\cite{eric}.
Notwithstanding the phenomenological successes of the naive quark
model, it suffers from the serious shortcoming of being unable to
account for the physics of chiral symmetry breaking. It therefore
fails, among other issues basically related to pion physics, to
constrain the treatment of strong decays by the underlying bound
state dynamics. This failure results, for example, in predictions
for couplings to the ground--state pseudoscalars which do not
satisfy PCAC or the Goldberger--Treiman
relations~\cite{Close:2005se}. Roughly speaking, quark models
consistent with chiral symmetry can be thought of being evolutions
of the naive quark model supplemented by the constraint of the
mass-gap equation, and therefore they should provide the correct
framework to study such strong decays. Here we shall show that such
type of models leads indeed to the GT relation for the pion
coupling between the opposite--parity states. This is a
non-trivial result, which can only follow from the simultaneous
use of a chiral invariant interaction together with a consistent
treatment of the pion and heavy--light meson dynamics. 

The paper
is organized as follows. In the following section we recapitulate
the chiral quark models and discuss the application to the
ground--state pseudoscalar and heavy-quark sectors. In
Section~\ref{GThl}, we present the proof of the
Goldberger--Treiman relation for the heavy--light mesons. Summary
and a brief discussion are given in Section~\ref{summary}.
Throughout the paper we concentrate on the $J^P=0^\mp$
heavy--light states, but the results can easily be generalized to
higher spins.

\section{Chiral quark models}
\label{cqm}

In this chapter, we give a short introduction to the chiral quark
model. The model is described by the Hamiltonian
\begin{eqnarray}
H& = & \int d^3x\psi^{\dag}({\x})\left(-i{\bm \alpha}
\cdot{\bm \bigtriangledown}+m\beta\right)\psi({\x}) \nonumber \\
&+ &  \frac12\int d^3 xd^3 y\;J^a_\mu({\x})K^{ab}_{\mu\nu}({\x}-{\y})J^b_\nu({\y}),
\label{H}
\end{eqnarray}
with the quark current--current
($J_{\mu}^a({\x})=\bar{\psi}({\x})\gamma_\mu\frac{\lambda^a}{2}\psi({\x})$)
interaction parameterized by the instantaneous confining kernel
$K^{ab}_{\mu\nu}({\x}-{\y})$. An important feature of the models
of the class (\ref{H}) is the remarkable robustness of their
predictions with respect to variations of the quark kernel.
Although quantitative results may vary for different kernels, the
qualitative picture described remains essentially the same. This
is specially true for those relations enforced by the mechanism of
chiral symmetry breaking  that should be independent of the
spatial details of the confining kernel provided it brings a
natural confinement scale (hereafter called $K_0$). The only
requirement is that it should be chirally symmetric. We illustrate
this nice feature by deriving the Goldberger--Treiman relation
connecting the pion coupling to heavy--light mesons to the mass
splitting between chiral doublets, for the simplest Lorentz
structure of the inter-quark interaction in the Hamiltonian
(\ref{H}) compatible with the requirements of chiral symmetry and
confinement
$K_{\mu\nu}^{ab}(\x-\y)=\delta^{ab}g_{\mu0}g_{\nu0}V(|\x-\y|)$~\cite{Orsay,Lisbon,ASES,Szczepaniak:1996tk},
for an arbitrary  confining kernel $V(r)$. 
We also note that analogous Hamiltonian with linearly rising potential 
considered in one time and
one spatial dimension reproduces the 't~Hooft model for 2D QCD \cite{tHooft} in the 
Coulomb (axial) gauge (see, for example, the
original paper \cite{BG} or the review paper \cite{2d} and references therein).

\begin{figure}[t]
\begin{center}
\includegraphics[width=3in]{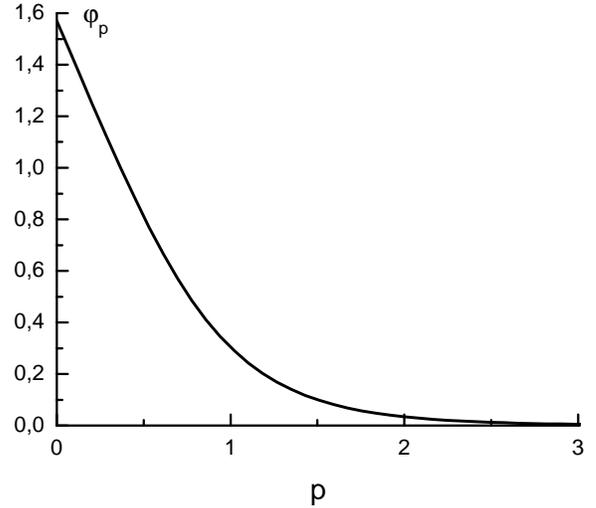}
\caption{A typical profile of the chiral angle --- solution to the
mass-gap equation. Momentum $p$ is measured in the units of the
potential strength $K_0$.}\label{vp}
\end{center}
\end{figure}

A standard way of proceeding with the investigation of the model
(\ref{H}) is to consider the self-interaction of quarks
separately, thus introducing the notion of the dressed quarks:
\begin{equation}
\psi(\x)=\sum_{\zeta=\uparrow,\downarrow}\int\frac{d^3p}{(2\pi)^3}
e^{i\p\x}[b_\zeta(\p)u_\zeta(\p)+d_\zeta^+(-\p)v_\zeta(-\p)],
\label{quarkfield2}
\end{equation}
where the quark amplitudes
\begin{equation}
\begin{array}{rcl}
u(\p)&=&\ds\frac{1}{\sqrt{2}}\left[\sqrt{1+s_p}+({\bm\alpha}
\cdot\hat{\p})\sqrt{1-s_p}\right]u_0(\p),\\[5mm]
v(-\p)&=&\ds\frac{1}{\sqrt{2}}\left[\sqrt{1+s_p}-(\bm{\alpha}
\cdot\hat{\p})\sqrt{1-s_p}\right]v_0(-\p)
\end{array}
\end{equation}
are parameterized with the help of the chiral angle $\vp_p$
\cite{Orsay,Orsay2,Lisbon}, $s_p=\sin\vp_p$, $c_p=\cos\vp_p$. It
is convenient to define the chiral angle varying in the range
$-\pi/2<\vp_p\leq\pi/2$, with the appropriate boundary conditions
$\vp(0)=\frac{\pi}{2}$, $\vp(p\to\infty)\to 0$ for the physical
vacuum. The equation which defines the profile of the chiral angle
--- the mass-gap equation --- follows from the requirement that
the quadratic part of the normally ordered Hamiltonian (\ref{H})
should be diagonal in terms of the dressed quark creation and
annihilation operators \cite{Lisbon}. The mass-gap equation then
takes the form:
\begin{equation}
mc_p-ps_p=\frac{C_F}{2}\int\frac{d^3k}{(2\pi)^3}V(\p-\k)[s_kc_p-xc_ks_p],
\label{mge}
\end{equation}
where $x=(\hat{\p}\cdot\hat{\k})$. In Fig.~\ref{vp}, we give a
typical profile of the chiral angle.

The dressed quark dispersive law can be evaluated then as
\begin{equation}
E_p = ms_p+p c_p +\frac{C_F}{2}\int \frac{d^3k}
{(2\pi)^3}V(\p-\k)\left[s_p s_k + x c_p c_k\right].
\end{equation}

We turn now to bound states of dressed quarks --- to the
quark--antiquark mesons. Each mesonic state in this model is
described with a two--component wave function, and the
bound--state equation acquires the form of a system of two coupled
equations. The details can be found in the works
\cite{Orsay2,Lisbon}, where the Bethe--Salpeter equation formalism
was developed or in Ref.~\cite{nr}, where a second Bogoliubov-like
transformation is performed over the Hamiltonian (\ref{H}) in
order to define explicitly the compound mesons creation and
annihilation operators. For the purpose of the present research,
it is sufficient to consider two particular cases of this general
mesonic bound--state equation.

\subsection{The case of the chiral pion}

In the chiral limit, the bound--state equation for the pion at
rest is known to be given by the mass-gap equation (\ref{mge})
with $m=0$ \cite{Lisbon,nr}.  In the random phase approximation
(RPA) both components of the pionic wave function coincide and are
simply $\sin\vp_p$ \cite{Orsay2,Lisbon,nr}. Beyond the chiral
limit, the pion state is given by $(\p_\pi=0)$:
\begin{eqnarray}
|\pi^i\rangle=\int\frac{d^3k}{(2\pi)^3}
\left[b^\dag(\k) \frac{\tau^i}{\sqrt{N_f}}
\frac{I_c}{\sqrt{N_C}}\frac{I_s}{\sqrt{2}}d^\dag(-\k) X(\k)\right.\nonumber \\
\\[-3mm]
+\left.d(-\k)\frac{\tau^i}{\sqrt{N_f}}
\frac{I_c}{\sqrt{N_C}}\frac{I_s}{\sqrt{2}}b(\k)Y(\k)\right]|0\rangle,\nonumber
\end{eqnarray}
with $|0\rangle$ being the broken BCS vacuum. $N_f$ and $N_C$ are,
respectively, the number of flavors and colors.

The normalized wave functions are given by
\begin{eqnarray}
X(\k)=\frac{\sqrt{N_C}\sqrt{N_s}\sqrt{N_f}}{2if_\pi}[s_k+m_\pi\Delta_k],\nonumber \\
\\[-3mm]
Y(\k)=\frac{\sqrt{N_C}\sqrt{N_s}\sqrt{N_f}}{2if_\pi}[s_k-m_\pi\Delta_k],\nonumber
\end{eqnarray}
with $\Delta_k$ satisfying
\begin{equation}
2E_k\Delta_k=s_k+C_F\int\frac{d^3q}{(2\pi)^3}V(\k-\q)[s_k s_q + x c_k c_q]\Delta_q.
\end{equation}

\subsection{The case of a heavy--light meson}

The case of a heavy--light meson, with one quark being infinitely
heavy, leads to considerable simplifications of the bound--state
equation in view of the fact that the negative energy--spin
component of the meson wave function, responsible for the
time--backward motion of the quark--antiquark pair, which is
non-negligible for the light-light system, goes to zero with the
bare mass increase of just one quark of the pair (heavy--light
systems), vanishing in the infinite mass limit of that quark. The
resulting equation takes the form of a Schr{\" o}dinger-like
equation \cite{parity}
\begin{eqnarray}
E_p\hat\Psi(\p)+\int\frac{d^3k}{(2\pi)^3}V(\p-\k)\left[C_pC_k+\right.\hspace*{1cm}\nonumber\\
\label{FW4} \\[-3mm]
\hspace*{1cm}\left.({\bm \sigma}\cdot \hat{\p})({\bm \sigma}\cdot \hat{\k})S_pS_k\right]
\hat \Psi(\k)=E\hat \Psi(\p)\nonumber,
\end{eqnarray}
with
\begin{eqnarray}
& & C_p=\frac{1}{\sqrt{2}}\left[\sqrt{1+c_p}+\sqrt{1-c_p}\right],\quad \nonumber \\
\\[-3mm]
& & S_p=\frac{1}{\sqrt{2}}\left[\sqrt{1+c_p}-\sqrt{1-c_p}\right].\nonumber
\end{eqnarray}

Here the wave function is written as a product of the orbital part
and the spin part, $\hat\Psi(k)= I/\sqrt{2}\cdot\Psi(k)$ for the
$J^P=0^-$ and $\hat\Psi'(k)= ({\bm
\sigma}\cdot\hat\k)/\sqrt{2}\cdot\Psi'(k)$ for the $J^P=0^+$
heavy--light meson, respectively. The conventional normalization of
the wave function (consistent with the relativistic normalization)
is
\begin{equation}
2M=\int\frac{d^3k}{(2\pi)^3}|\Psi(\k)|^2=\int\frac{d^3k}{(2\pi)^3}|\Psi'(\k)|^2,
\end{equation}
where, in the heavy--quark limit, $M$ approximates the mass of the
heavy quark, and the mass of the heavy--light meson is given by
$M+E$.

\section{Goldberger--Treiman relation for heavy--light mesons}
\label{GThl}

In this chapter, we focus on the details of the pion interaction
with heavy--light mesons. Consider the hadro\-nic process which
involves the chiral pion and the two heavy--light mesons for
which, for the sake of transparency, we consider the ${\bar D}(J^P=0^+)$ and
the ${\bar D}'(J^P=0^-)$ mesons with the wave functions $\hat\Psi(k)$ and
$\hat\Psi'(k)$, respectively.

\subsection{The macroscopic derivation of the Goldberger--Treiman relation in QCD}

Let us consider the transition ${\bar D}\to {\bar D}'\pi$. Then, using the relation
\begin{equation}
\langle 0|A^a_\mu(0)|\pi^b(\q)\rangle=if_\pi q_\mu\delta^{ab},
\end{equation}
we introduce the $n n'\pi$ coupling $g_{nn'\pi}$ through the relation
\begin{equation}
\langle n'|A^a_\mu|n\rangle=\langle n'|A_\mu^a|n\rangle_{\mbox{nonpion}}-
\frac{2Mq_\mu f_\pi g_{nn'\pi}}{q^2-m_\pi^2+i\epsilon} D^{'\dag}\tau^aD,
\label{A}
\end{equation}
where the subscript ``nonpion" denotes the contribution free of the pion pole.
Here $D'$, $D$ are unit isospin doublets representing the flavor of the heavy--light $n'$ and $n$ state, respectively. 

On the other hand, the full matrix element of the axial current, corresponding to the {\em l.h.s.} of Eq.~(\ref{A}) can be written as:
\begin{eqnarray}
\langle n'|A^a_\mu|n\rangle &=& \left[(P'_\mu+P_\mu)G_A(q^2)\right.\nonumber \\
\\[-3mm]
&-& \left.(P'_\mu-P_\mu)G_S(q^2)\right]D'\frac{\tau^a}{2}D,\nonumber
\end{eqnarray}
with $P'=p_{n'}$, $P=p_n$ $q=P'-P$. To leading order in the heavy mass, conservation of the
axial current in the chiral limit  then demands
\begin{equation}
2M(E'-E)G_A- q^2G_S=0.
\label{GT}
\end{equation}

Therefore, from Eq.~(\ref{A}) one can see that as $q^2\to 0$ it is
possible to have $G_A(0)\ne 0$, if $G_S$ is identified with the
pion pole contribution,
\begin{eqnarray}
\lim_{q^2\to 0}G_S(q^2)\to\frac{4Mf_\pi g_{n n'\pi}}{q^2},
\end{eqnarray}
and $G_A$ remains finite in the limit $q^2\to 0$. The relation (\ref{GT}) then reads:
\begin{equation}
\frac12(E'-E)G_A=f_\pi g_{n n'\pi}
\label{GTh}
\end{equation}
and is the sought Goldberger--Treiman relation for the pion coupled to heavy--light mesons.

\subsection{The microscopic derivation of the Goldberger--Treiman relation in the chiral quark model}

We now turn to the main subject of this paper and demonstrate how
the Goldberger--Treiman relation (\ref{GTh}) emerges
microscopically in the chiral quark model.

We are interested in matrix elements of the type $\langle
n'|O|n\rangle$ where, in particular, $O$ is the isovector axial current operator
$A^a_\mu(0)=\bar{\psi}(0)\gamma_\mu\frac{\tau^a}{2}\gamma_5\psi(0)$, whereas
$|n\rangle$ and $|n'\rangle$ are the true
eigenstates of the QCD Hamiltonian $H_{QCD}$ which, after
integrating out gluon degrees of freedom, we approximate by the Hamiltonian of the chiral quark model (\ref{H}).
Formally, the Hamiltonian (\ref{H}) is represented in the Fock space of
single hadrons built on top of the RPA vacuum. It is convenient to
split the Hamiltonian in two parts,
\begin{equation}
H_{QCD}=(H_{QCD}-V)+V\equiv{\tilde H}+V \approx H,
\end{equation}
where $V$ describes interactions with pions and $\tilde{H}$ is
assumed to produce the bare spectrum of hadronic states except for
mass shifts. 
The bare part ${\tilde H}$ is then identified with those parts of the Hamiltonian
(\ref{H}), which, in the dressed quark basis (\ref{quarkfield2}),
do not create (or annihilate) single light quark--antiquark pairs.
The latter are put into $V$ and lead to non-vanishing matrix
elements $V_{nn'\pi}$ between the opposite--parity 
meson states with one additional pion. 
Bearing in mind the low--energy limit, we thus assume
that the spectrum can be well approximated by the bare spectrum
(in the large-$N_C$ limit) and the only relevant residual
interactions are those with pions. In the interaction picture, the
matrix elements of interest can be written as
\begin{equation}
\langle n'|O|n\rangle=\langle {\tilde n}'|U(+\infty,0)OU(0,-\infty)|\tilde{n}\rangle,
\end{equation}
where the states on the {\em r.h.s.} marked with tildes are
eigenstates of the bare, pion coupling free, Hamiltonian, ${\tilde
H}$ and $U$ is the evolution operator,
\begin{equation}
U(t_f,t_i)=T\exp
\left[-i\int_{t_i}^{t_f}dt\left(e^{i\tilde{H}t}Ve^{-|t|\epsilon}e^{-i\tilde{H}t}\right)\right].
\end{equation}
To leading order in the pion emission/absorption (that is, in $V$)  we obtain
\begin{eqnarray}
\langle n'|O|n\rangle=\langle{\tilde n}'|O|\tilde{n}\rangle&+&
\sum_m\frac{\langle {\tilde n}'|O|{\tilde m}
\rangle\langle{\tilde m}|V|{\tilde n}\rangle}{E_n-E_m+i\epsilon}\nonumber \\
\label{time} \\[-3mm]
&+&\sum_m\frac{\langle{\tilde n}'|V|\tilde{m}
\rangle\langle{\tilde m}|O|{\tilde n}\rangle}{E_{n'}-E_m+i\epsilon}.\nonumber
\end{eqnarray}

The intermediate states always contain an extra pion. Now, because
we are interested in the case $O\equiv A^a_\mu(0)$, that is, in matrix
elements of the type $\langle 0|A^a_\mu(0)|\pi^b(\k)\rangle$, we
can, for the case of the heavy--light $n\to n'\pi$ transitions,
saturate $\sum_m |{\tilde m}\rangle\langle{\tilde m}|$ as follows, 
($N=n, \mbox{ or } n'$) 
\begin{equation}
\sum_m |{\tilde m}\rangle\langle{\tilde
m}|\rightarrow |{\tilde N}\rangle\langle {\tilde N}|\sum_a\int\frac{d^3k}
{2\omega(\k)(2\pi)^3}|\pi^a(\k)\rangle\langle\pi^a(\k)|.
\label{pi}
\end{equation}

For the pion interacting with heavy--light mesons, the coupling constant $g_{nn'\pi}$ can be introduced then as
\begin{equation}
\langle {\bar D}'\pi^a |V|{\bar D} \rangle=2M i g_{nn'\pi}D^{'\dag} \tau^a D(2\pi)^3\delta^{(3)}({\bf P}'+\p_\pi-{\bf P}),
\label{Vnn}
\end{equation}
and therefore, in order to derive the Goldberger--Treiman relation, we are to evaluate microscopically the matrix element on the 
{\em l.h.s.} of Eq.~(\ref{Vnn}). 

\begin{figure}[t]
\begin{center}
\includegraphics[width=8.5cm]{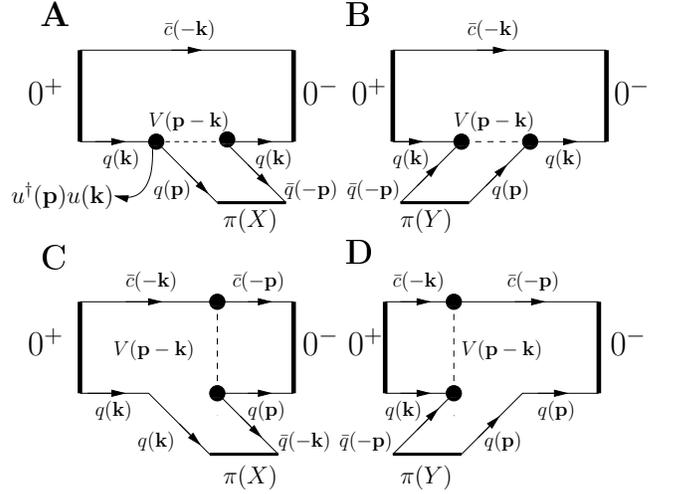}
\end{center}
\caption{The four diagrams contributing to $\langle {\bar D}'\pi |V|{\bar D}\rangle$.}\label{ABCD}
\end{figure}

Obviously, the interaction potential $V$, when written in the quark basis, is given
by the four diagrams shown in Fig.~\ref{ABCD}. In these diagrams
$\bar{c}$ stands for the charm antiquark. The black dots correspond
to the appropriate spinor vertices --- see Ref.~\cite{Lisbon}. As
an example, we depict in diagram A one such vertex:
$u^\dag(\p)u(\k)=\frac{1}{2}[\sqrt{1+s_p}\sqrt{1+s_k}+({\bm
\sigma}\cdot\hat{\p}) ({\bm
\sigma}\cdot\hat{\k})\sqrt{1-s_p}\sqrt{1-s_k}]$. As it was
discussed before, the pion Salpeter amplitude has two components:
the positive energy--spin component $X$ and the negative
energy--spin component $Y$. The four diagrams of Fig.~\ref{ABCD}
can be readily evaluated using the rules of Ref.~\cite{Lisbon}.
For the chiral pion at rest, the diagrams A and B can be shown
explicitly to cancel against each other so that one is left only
with the diagrams C and D. This should not come as a surprise,
once, in the chiral limit, pions at rest decouple from quarks. We
use the notations $C_X$ and $D_Y$ for the amplitudes corresponding
to the matrix elements of $V$ shown in Fig.~\ref{ABCD} (diagrams C
and D) with the interaction attached to the heavy quark. Thus we
have
\begin{equation}
\langle {\bar D}'\pi^a |V|{\bar D} \rangle=2M\left[C^a_X+D^a_Y\right](2\pi)^3\delta^{(3)}({\bf P}'+\p_\pi-{\bf P})
\end{equation}
with the amplitudes $C_X$ and $D_Y$ calculated explicitly to be
\begin{widetext}
\begin{eqnarray}
C^a_X=\frac{D'\tau^a D}{2\sqrt{2} M}\frac{1}{\sqrt{N_f}}\frac{1}{\sqrt{N_C}}\frac{C_F}{2}
\int\frac{d^3k}{(2\pi)^3}\frac{d^3q}{(2\pi)^3}
\Psi'(\q)V(\q-\k)X(\k)\Psi(\k)[\sqrt{1-s_q}\sqrt{1+s_k}-x\sqrt{1+s_q}\sqrt{1-s_k}],\nonumber \\
\\[-3mm]
D^a_Y=\frac{D'\tau^a D}{2\sqrt{2} M}\frac{1}{\sqrt{N_f}}\frac{1}{\sqrt{N_C}}\frac{C_F}{2}
\int\frac{d^3k}{(2\pi)^3}\frac{d^3q}{(2\pi)^3}
\Psi'(\q)V(\q-\k)Y(\q)\Psi(\k)[\sqrt{1 - s_q}\sqrt{1 + s_k}-x\sqrt{1+s_q}\sqrt{1-s_k}]\nonumber.
\end{eqnarray}
The high momentum (UV)  behavior of integrals is determined by the short distance behavior of the kernel and wave functions. For purely confining  kernels wave functions are soft and integrals are converging without need for renormalization. The short distance behavior of the kernel depends on integrating out gluons. In the approximation that keeps transverse gluons out from the Fock space the self-consistent solution the Dyson equation for the kernel leads to UV Coulomb potential with a running coupling softer then $\sim~1/\log(q^2)$ and leads to finite UV integrals. This is discussed, for example in~\cite{ASES,Szczepaniak:2002ir}.  In the IR all integrals are finite for long range confining potentials for color singlet matrix elements  as discussed, for example  in~\cite{Orsay,Orsay2,Lisbon}. Adding the two amplitudes together and using the definition of $g_{{\bar D}{\bar D}'\pi}$,
to leading order in $m_\pi$, we get:
\begin{equation}
f_\pi g_{{\bar D}{\bar D}'\pi}=\frac{C_F}{2}\frac{1}{2M}\int\frac{d^3k}{(2\pi)^3}\frac{d^3q}{(2\pi)^3}
\Psi'(\q)V(\q-\k)\frac{s_k + s_q}{2}\Psi(\k)[\sqrt{1-s_q}\sqrt{1+s_k}-x\sqrt{1+s_q}\sqrt{1-s_k}].
\label{gp}
\end{equation}

The non-pion contribution to the axial charge is computed from the
expectation value of the time component of the axial current
between the heavy--light meson states:
\begin{equation}
G_A=\frac{1}{2M}\int\frac{d^3k}{(2\pi)^3}\Psi'(\k)c_k\Psi(\k).
\end{equation}

Finally, integrating the Schr\"odinger-like (\ref{FW4}) for
$\Psi(\k)$ with $\Psi'(\k)c_k$ and, consequently, a similar
equation for $\Psi'(\k)$ with $\Psi(\k)c_k$, we find:
\begin{eqnarray}
\frac12(E'-E)G_A=\frac{C_F}{4}\frac{1}{2M}\int\frac{d^3k}{(2\pi)^3}\frac{d^3q}{(2\pi)^3}
\Psi'(\q)V(\q-\k)[c_q(\sqrt{1+s_q}\sqrt{1+s_k}+x\sqrt{1-s_q}\sqrt{1-s_k})\nonumber \\
\\[-3mm]
-c_k(x\sqrt{1+s_q}\sqrt{1+s_k}+\sqrt{1-s_k}\sqrt{1-s_q})]\Psi(\k)\nonumber \\
\end{eqnarray}
or, after simple algebraic transformations:
\begin{equation}
\frac12(E'-E)G_A=\frac{C_F}{2}\frac{1}{2M}\int\frac{d^3k}{(2\pi)^3}\frac{d^3q}{(2\pi)^3}
\Psi'(\q)V(\q-\k)\frac{s_k + s_q}{2}\Psi(\k)[\sqrt{1-s_q}\sqrt{1+s_k}-x\sqrt{1+s_q}\sqrt{1-s_k}].
\label{EEG}
\end{equation}
\end{widetext}

Comparing Eqs.~(\ref{gp}) and (\ref{EEG}) we arrive immediately at
the Goldberger--Treiman relation (\ref{GTh}).

\section{Discussion}
\label{summary}

In this paper, we proved the validity of the Goldberger--Treiman
relation (\ref{GTh}) in the chiral quark model. 
We assumed that the model Hamiltonian represents the fourth-compenet of a four-vector in a particular frame, the one  in which the heavy mesons are at rest and can verify relations between matrix elements involving fourth  component of vectors.  Several important
conclusions should be drawn from the presented consideration.
First of all, it should be noted that had we dropped the $Y$ piece
of the pion wave function (as in naive quark models) we would have
been not only violating the Goldberger--Treiman relation by $50\%$
as should be expected, but also the two diagrams, A and B (with
the interaction coupled solely to the light quarks), would have
now survived to give a result incompatible with the $G_A$
contribution. This is yet another manifestation of the Goldstone
nature of the chiral pion which finds its natural implication in
the chiral model used in this work. Secondly, the
Goldberger--Treiman relation can only be realized by the Salpeter
solutions $\Psi$ and $\Psi'$ of Ref.~\cite{parity}, which in turn
contain the physics of progressive effective restoration of chiral
symmetry for higher and higher excitations. Moreover, with the
Goldberger--Treiman relation (\ref{GTh}), we are in a position to
go even further and to conclude that the pion coupling to excited
mesons decreases as the excitation number of the heavy--light
mesons increases --- according to the Goldberger--Treiman relation
(\ref{GTh}) with $\Delta M=M'-M\too 0$ and $G_A\too 1$. This gives
an explicit pattern of the Goldstone boson decoupling from excited
hadrons. Such a scenario was discussed in a recent paper \cite{NG},
where the decrease of the pion coupling to excited hadrons was anticipated 
as a consequence of the chiral pion decoupling from dressed quarks. In the present work, this result
comes out naturally in the microscopic derivation of the
Goldberger--Treiman relation and, at the same time, an explicit
expression for the ${\bar D}{\bar D}'\pi$ coupling constant is found. 

\section{Acknowledgment}

Work of A.N. was supported by the Federal Agency for Atomic Energy of Russian Federation, 
by the Federal Programme of the Russian Ministry of Industry, Science, and Technology No.
40.052.1.1.1112, by the Russian Governmental Agreement N 02.434. 11.7091, and by grants 
RFFI-05-02-04012-NNIOa, DFG-436 RUS 113/820/0-1(R), NSh-843.2006.2.  
A.P.S. would like to acknowledge support of the US DOE via the contract DE-FG0287ER40365 and to thank the support
of the CFIF of the Instituto Superior T\'ecnico where part of this work was completed.

\end{document}